\documentclass[aps,prb,amsmath,twocolumn]{revtex4}
\usepackage{graphicx}
\usepackage{bm}

\begin{document}
\title{Transport of interacting electrons in arrays of quantum dots and diffusive wires}

\author{Dmitri S. Golubev$^{1,3}$ and
Andrei D. Zaikin$^{2,3}$}
\affiliation{$^1$Institut f\"ur Theoretische Festk\"orperphysik,
Universit\"at Karlsruhe, 76128 Karlsruhe, Germany \\
$^{2}$Forschungszentrum Karlsruhe, Institut f\"ur Nanotechnologie,
76021, Karlsruhe, Germany\\
$^{3}$I.E. Tamm Department of Theoretical Physics, P.N.
Lebedev Physics Institute, 119991 Moscow, Russia}

\begin{abstract}
We develop a detailed theoretical investigation of the effect of
Coulomb interaction on electron transport
in arrays of chaotic quantum dots and diffusive metallic wires. 
Employing the real time path integral technique we formulate 
a new Langevin-type of approach which exploits a direct
relation between shot noise and interaction effects in mesoscopic conductors.
With the aid of this approach we establish a general expression
for the Fano factor of 1D quantum dot arrays and derive a complete
formula for the interaction correction to the current
which embraces all perturbative results previously obtained
for various quasi-0D and quasi-1D disordered conductors and
extends these results to yet unexplored regimes.

\end{abstract}
\maketitle

\section{Introduction}
Recently a profound relation was elucidated
\cite{GZ00,GGZ02,Naz2,BN} between full counting statistics (FCS)
\cite{LLL} and electron-electron interaction effects in coherent
mesoscopic conductors. In particular, it was demonstrated
\cite{GZ00} that the leading interaction correction to the current
through such conductors is determined by the second cumulant of
the current operator ${\cal S}_2$, i.e. by the power spectrum of
the shot noise \cite{bb}. The interaction correction to the shot
noise ${\cal S}_2$ was in turn found \cite{GGZ02} proportional to
the third cumulant of the current operator ${\cal S}_3$. Even more
generally, one can show \cite{Naz2,BN} that the lowest order
interaction correction to the $n$-th current cumulant ${\cal S}_{n}$ is
determined by ${\cal S}_{n+1}$ for all values of $n$. Since the current
cumulants in the absence of interactions can be conveniently
analyzed within the FCS formalism \cite{LLL}, the above
observations provide a great deal of information about the effect
of electron-electron interactions as well.

In order to investigate the influence of interactions on higher
current cumulants it is in general  necessary to employ a
complete expression for the effective action of a coherent
scatterer \cite{GGZ02,Naz2,BN}.  At the same time the results
\cite{GZ00} for the {\it first} cumulant, i.e. the relation
between the leading interaction correction to the current and the
shot noise can be understood already within a simple and
transparent theoretical framework of quasiclassical Langevin
equations. In the case of a single coherent scatterer shunted by
some linear Ohmic resistor $R_S$ these equations take a remarkably
simple form
\begin{eqnarray}
C\frac{\ddot\varphi}{e}+\frac{1}{R}\frac{\dot\varphi}{e}&=&I(t)+\xi(t),
\nonumber\\
\frac{1}{R_S}\left(V_x-\frac{\dot\varphi}{e}\right)&=&I(t)+\xi_S(t).
\label{single}
\end{eqnarray}
Here $C$ is the scatterer capacitance, $\dot\varphi /e=V$ is the
fluctuating voltage across the scatterer and $V_x$ is the total
voltage applied to the system ``scatterer+shunt''. As usually, one
describes the scatterer by a set of conducting channels with
transmissions $T_k$. The scatterer conductance is then defined by
means of the standard Landauer formula
\begin{equation}
\frac{1}{R}=\frac{e^2}{\pi}\sum_k T_k,
\end{equation}
 $\xi(t)$ is the noise of the scatterer, characterized by the correlator
\begin{eqnarray}
\langle\xi(t_1)\xi(t_2)\rangle &=& \frac{1-\beta +\beta \cos[\varphi(t_1)-\varphi(t_2)]}{R}
\nonumber\\ &&\times\,
\int\frac{d\omega}{2\pi}\,\omega\coth\frac{\omega}{2T}\,{\rm e}^{-i\omega(t_1-t_2)},
\label{noise1}
\end{eqnarray}
where
\begin{equation}
\beta =\frac{\sum_k T_k(1-T_k)}{\sum_k T_k}
\end{equation}
is the Fano factor,
and $\xi_S(t)$ is the equilibrium noise of the shunt with the correlator
\begin{equation}
\langle\xi_S(t_1)\xi_S(t_2)\rangle=\frac{1}{R_S}\int\frac{d\omega}{2\pi}\,
\omega\coth\frac{\omega}{2T}\,{\rm e}^{-i\omega(t_1-t_2)}.
\label{xis}
\end{equation}

The whole approach based on Eqs. (\ref{single}-\ref{xis}) is
applicable either at sufficiently high energies or, more
importantly, if at least one of the two dimensionless
conductances, $g=2\pi /e^2R$ and/or $g_S=2\pi /e^2R_S$, remains
large. The above Langevin equations make the relation between the
interaction correction to the current and the shot noise
completely transparent demonstrating that the former can be
derived just if one accounts for the noise contribution in the
equation describing the balance of currents across the scatterer.

The above simple approach may hold only for relatively compact
scatterers, in which case the description of interaction effects
with the aid of the uniform in space fluctuating field $\varphi$
is sufficient. For spatially extended conductors the coordinate
dependence of this field cannot anymore be disregarded and the
whole analysis needs to be modified. This modification is trivial
if one considers an array of scatterers connected between each
other by relatively big metallic grains. Assuming that the
electron distribution function in each of these grains remains in
equilibrium one can describe the $n$-th scatterer
by its own fluctuating field $\varphi_n$ which obeys the same set
of Langevin equations (\ref{single}), (\ref{noise1}). For arrays
of tunnel junctions this approach was employed in Ref.
\onlinecite{many3}. The corresponding generalization of the
results \cite{many3} to the case of arbitrary scatterers just
requires modification of the Fano factor in the noise correlator
(\ref{noise1}).

The condition of local equilibrium inside metallic grains implies
that the corresponding electron dwell time $\tau_D$ between two
adjacent scatterers should be much longer than the inelastic relaxation
time $\tau_{\rm in}$. If this condition is violated, the electron
distribution function is driven out of equilibrium and the whole
consideration becomes more complicated. In the case of a quantum
dot formed by two arbitrary scatterers the latter situation was
analyzed in details in Ref. \onlinecite{GZ03} and also in Refs.
\onlinecite{SR,BN} for the case of chaotic dots. In all these
works it was demonstrated that in the limit of large conductances
and in the voltage biased regime the interaction correction to the
conductance tends to saturate at temperatures/voltages below
$1/\tau_D$. This implies that for finite values of $\tau_D$ highly
conducting quantum dots should show metallic behavior down to zero
temperature.

It is important to emphasize that this observation holds only
provided the voltage source is attached {\it directly} to the
quantum dot, i.e. the external impedance is equal to zero. For
non-zero external impedances voltage fluctuations lift the
conductance saturation, and the amplitude of the interaction
correction keeps increasing with decreasing $T$ even at
temperatures well below $1/\tau_D$. In this regime the interaction
correction was found \cite{GZ03} to scale linearly with the total
Fano factor of the quantum dot and to depend logarithmically on
temperature/voltage for sufficiently large external impedances or
if this impedance is purely Ohmic.

For similar reasons no saturation of the interaction correction at
energies below $1/\tau_D$ should be expected for chains and arrays
of quantum dots. Recently this situation was analyzed
diagrammatically \cite{many4} in the case of granular tunnel
junction arrays. Indeed, it was found that the interaction
correction increases with decreasing temperature both above and
below the inverse dwell time in individual grains. At $T <
1/\tau_D$ the authors \cite{many4} recovered exactly the same
expression for the interaction correction as that known in the
case of diffusive conductors \cite{AA}. This equivalence is by no
means surprising if one bears in mind the fundamental relation
between the interaction correction and the shot noise on one hand,
and the results \cite{Jal,Sukh} on the other hand, which demonstrate
that the shot noise of a sufficiently long array of tunnel
junctions is equivalent to that of a diffusive wire. Extending
these arguments to arbitrary scatterers, with the aid of the 
results \cite{Sukh} one can
anticipate that at sufficiently low energies ($\lesssim 1/\tau_D$)
and large spatial scales the interaction correction should be described
universally for any array of quantum dots and ultimately for {\it
any} mesoscopic conductor in the metallic regime. This
universality will indeed be demonstrated below.

The main goal of the present paper is to generalize the simple
Langevin equation approach \cite{many3} to situations in which
relaxation of the electron distribution function occurs at much
longer time scales as compared to the electron dwell time between
two adjacent scatterers $\tau_D\ll\tau_{\rm in}.$ Although the
distribution function may significantly deviate from the Fermi
function, it is possible to account for these
deviations within the (generalized) Langevin equation analysis and
to formulate a closed set of equations which fully determine the
interaction correction to the $I-V$ curve of disordered
conductors.

The structure of the paper is as follows. In Sec. II we will
specify the model of a disordered metallic conductor and present a
phenomenological derivation of the basic Langevin equations for
our problem. This derivation will be carried out with the aid of
simple and transparent physical arguments which make the whole
approach easy to understand without going into technical details.
A more advanced analysis employing the effective action technique
will be described in Sec. III. This analysis provides rigorous
justification for our phenomenological derivation and allows to
illustrate a useful relation between our technique and the
classical Boltzmann-Langevin approach \cite{bb}. In Sec. IV we
will probe our Langevin technique by explicitly deriving the shot
noise spectrum and the Fano factor for arrays of 
chaotic quantum dots in the absence of interactions. The remainder
of the paper will be devoted to the analysis of the leading
interaction correction to the current in arrays of quantum dots
and mesoscopic diffusive wires. In Sec. V we will derive the
general expression for this correction which then will be applied
to homogeneous arrays of quantum dots in Sec. VI. In the latter
case we will establish a complete analytic form of the interaction
correction and present the corresponding simplified expressions in
a number of important limits. Our general formula, Eq.
(\ref{dIexact}), embraces all previous results
\cite{GZ00,BN,many3,GZ03,SR,many4,AA,GG} obtained in various types
of quasi-0D and quasi-1D disordered conductors, allows to
establish a transparent relation between these results and to
extend them to yet unexplored regimes. A brief analysis of an
additional effect of external leads will be presented in Sec. VII.
We will then discuss our results and conclude the paper in Sec. VIII.

\section{The model and phenomenological analysis}

We shall consider a chain of $N-1$ quantum dots as it is shown in
Fig. 1. Each dot can be viewed as an island in-between two
scatterers/barriers which in turn connect adjacent quantum dots.
Electrons can enter the dot through one of the barriers, spend
some time there propagating between the barriers, possibly being
scattered at the barriers, outer walls or otherwise, and finally
leave the dot through another barrier. In what follows we will
adopt the model of {\it chaotic} quantum dots.

Each of the $N$ barriers will be described by its Landauer
conductance $1/R_n=(e^2/\pi)\sum_k T_k^{(n)}$, capacitance $C_n$
and Fano factor $\beta_n=\sum_k T_k^{(n)}(1-T_k^{(n)}) /\sum_k
T_k^{(n)}$, where $T^{(n)}_k$ is the transmission of the $k$-th
conducting mode in the $n$-th barrier. We also define
dimensionless conductances of the scatterers $g_n=2\pi/ e^2R_n$.
In what follows we will assume that each scatterer has many
conducting channels and that its dimensionless conductance is
large $g_n\gg 1$. The $n$-th dot will be characterized by the mean
level spacing $\delta_n=1/N_0{\cal V}_n$, where ${\cal V}_n$ is
the dot volume and $N_0$ is the density of states at the Fermi
level. For the sake of generality we will also assume that each
dot has an additional capacitance to the ground $C_{gn}$. Finally,
the first and the last scatterers are connected to two big
metallic reservoirs which in turn are connected to the voltage
source via external leads with an Ohmic resistance $R_S$.

\begin{figure}
\begin{center}
\includegraphics[width=9cm]{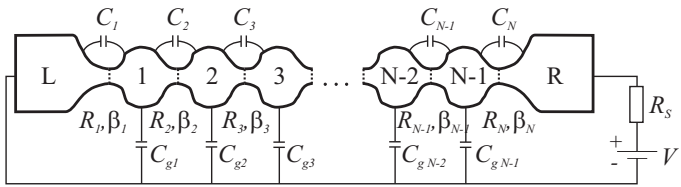}
\end{center}
\caption{1D array of chaotic quantum dots. The array consists of
$N-1$ dots and $N$ barriers. The $n$-th dot is characterized by
mean level spacing $\delta_n$ and gate capacitance $C_{gn}$. The
$n$-th barrier is described by its Landauer conductance $1/R_n$,
capacitance $C_n$ and Fano factor $\beta_n$. The array is placed
in-between two big metallic reservoirs which are connected to the
voltage source via Ohmic resistor $R_S$.}
\end{figure}

An important assumption concerns the spatial dependence of
fluctuating voltages in our system. Similarly to Ref.
\onlinecite{GZ03} we will allow for voltage drops $V_n(t)$ only across
the barriers, while inside the dots voltages can depend
arbitrarily on time but not on the spatial coordinates.
This assumption is usually well satisfied for metallic dots
considered here. In the leads the voltage fields are assumed to
vary slowly in space. In the course of our derivation we will
essentially neglect all mechanisms of inelastic relaxation which
are not related to electron-electron interactions. We will also
disregard weak localization effects which can be easily
suppressed, e.g., by externally applied magnetic field.

We will now proceed with our phenomenological analysis of the
above model.

\subsection{Noise correlator}

As a first step we will specify the general expression for the
noise correlator needed for our derivation. Let us assume that the
electron distribution function $f_n(E)$ in the $n$-th dot does not
depend on time for all $n$ but may deviate arbitrarily from the
Fermi function. Below we also assume that the electron energies
are measured with respect to the unique reference energy for the
whole array. In this case the noise of the $n-$th scatterer ${\cal
S}_n(t_1,t_2)=\langle\xi_n(t_1)\xi_n(t_2)\rangle$ takes the form
\cite{bb}
\begin{eqnarray}
{\cal S}_n(t_1,t_2) =\frac{1}{2R_n}\int\frac{d\omega}{2\pi}\int
dE\,{\rm e}^{-i\omega(t_1-t_2)}
\hspace{1.5cm} 
\nonumber\\  \times
\big\{
\beta_n
\big[f_{n-1}(E)h_n(E+\omega) +f_n(E+\omega)h_{n-1}(E) 
\nonumber\\
 +f_{n}(E)h_{n-1}(E+\omega)+f_{n-1}(E+\omega)h_n(E) \big]
\nonumber\\  +(1-\beta_n)\big[f_{n-1}(E)h_{n-1}(E+\omega)
\nonumber\\  +f_{n-1}(E+\omega)h_{n-1}(E) +f_n(E)h_n(E+\omega)
\nonumber\\  +f_n(E+\omega)h_n(E) \big]\big\}, \label{noise2}
\end{eqnarray}
where $h_n(E)=1-f_n(E).$
Let us define the function
\begin{equation}
G_n(t_1,t_2)=\int\frac{dE}{2\pi}\,{\rm e}^{-iE(t_1-t_2)}\,\big(1-2f_n(E)\big).
\label{gn1}
\end{equation}
In stationary situations this function depends only on the time
difference $t_1-t_2$ and it also obeys the condition
$G_n(t_2,t_1)=G_n^*(t_1,t_2).$ In equilibrium the distribution function
equals to the Fermi function
\begin{equation}
f_F(E)=\frac{1}{1+\exp\left(\frac{E}{T}\right)}. \label{fF}
\end{equation}
Substituting this function into Eq. (\ref{gn1}) one finds
$G_n(t_1,t_2)={-iT}\,{\rm Re}\,\big[1/{\sinh\pi
T(t_1-t_2+i\epsilon)}\big],$ where $\epsilon$ is an infinitesimal
positive constant. At $t_1\to t_2$ one gets $G_n(t_1,t_2)\to
{-i}\,{\rm Re}\,\big[1/\pi
  (t_1-t_2+i\epsilon)\big]$.
This analytical property turns out to be general, i.e. it equally
applies both to equilibrium and non-equilibrium situations.
Expressing Eq. (\ref{noise2}) via $G_n(t_1,t_2)$, we obtain
\begin{eqnarray}
\label{noise3}
{\cal S}_n(t_1,t_2) = \frac{1}{\pi
R_n}\frac{\epsilon^2}{((t_1-t_2)^2+\epsilon^2)^2} \hspace{3cm}
\\
-\,\frac{\pi \beta_n}{2R_n}\,
\big[G_{n-1}(t_1,t_2)G_n(t_2,t_1)
+G_{n}(t_1,t_2)G_{n-1}(t_2,t_1)\big]
\nonumber\\
-\,\frac{\pi (1-\beta_n)}{2R_n}
\big[|G_{n-1}(t_1,t_2)|^2+|G_{n}(t_1,t_2)|^2\big].
\nonumber
\end{eqnarray}
Although the formula (\ref{noise3}) has been derived
under the assumption that the distribution function does not
depend on time, we will show later that it remains valid also in
non-stationary situations. In the latter case the function
$G_n(t_1,t_2)$ can be understood as the Keldysh component of the
quasiclassical Usadel Green function.

\subsection{Kinetic equation}

Our next step is to derive the kinetic equation for the function
$G_n(t_1,t_2).$ For simplicity  we again start from the stationary
situation, in which case fluctuations of  voltages in our system
can be neglected. In what follows we will assume that both
$f_n(E)$ and $G_n(t_1,t_2)$ do not depend on coordinates inside
the $n-$th dot. The total number of electrons with energies in the
interval $[E,E+dE]$ in the $n$-th dot is $2N_0{\cal V}_nf_n(E)dE,$
where the factor 2 accounts for spin degeneracy. This number may
change in time only if electrons leave and/or enter the dot trough
the left ($n-$th) and the right ($n+1-$th) barriers. One finds
\begin{equation}
\frac{2dE}{\delta_n}\frac{\partial f_n(t,E)}{\partial
t}=J_{n}^{in}-J_{n}^{out}.
\end{equation}
The escape rate from the $n$-th dot and the transition rate to
this dot -- both through the $n-$th scatterer -- are respectively
\begin{equation}
\Gamma_{n-1,n}=g_n\delta_n/4\pi,\;\;\;
\Gamma_{n,n-1}=g_n\delta_{n-1}/4\pi .
\end{equation}
Then for $J_n^{out}$ one obtains
$$J_n^{out}=2(\Gamma_{n+1,n}+\Gamma_{n-1,n})f_n(E)dE/\delta_n,$$
and similarly for $J_n^{in}$. Combining the above simple
expressions we arrive at the kinetic equation
\begin{equation}
\frac{4\pi}{\delta_n}\frac{\partial f_n}{\partial
t}=-(g_n+g_{n+1})f_n+g_nf_{n-1}+g_{n+1}f_{n+1}, \label{kin1}
\end{equation}
where $1 \leq n \leq N-1$. The boundary conditions to this kinetic
equation are set by the requirement that the distribution
functions in the left and the right reservoirs, $f_0$ and $f_N$, are
equal to the Fermi function, i.e.
\begin{equation}
f_0(E)=f_F(E),\;\;\;\; f_N(E)=f_F(E-eV). \label{boundary1}
\end{equation}
Here and below $V$ is the total voltage applied to the array. We
note that the function $1-2f_n$ also satisfies Eq. (\ref{kin1}).

The kinetic equation for the function $G_n(t_1,t_2)$ can be
obtained from Eq. (\ref{kin1}) if we identify $t=(t_1+t_2)/2,$
introduce $s=t_1-t_2$ and make the Fourier transform of
(\ref{kin1}) by taking the integral $\int\frac{dE}{2\pi}\,{\rm
e}^{-iEs}(\dots )$. Then we obtain
\begin{eqnarray}
\frac{4\pi}{\delta_n}\frac{\partial G_n(t,s)}{\partial t}&=&-(g_n+g_{n+1})G_n(t,s)+g_nG_{n-1}(t,s)
\nonumber\\ &&
+\,g_{n+1}G_{n+1}(t,s).
\label{kin2}
\end{eqnarray}
As we have already pointed out Eq. (\ref{kin2}) applies only in
stationary situations. A proper generalization of this equation
for non-stationary cases can be achieved with the aid of general
gauge invariance arguments which yield
\begin{eqnarray}
\frac{4\pi}{\delta_n}\left(\frac{\partial }{\partial t_1}+\frac{\partial }{\partial t_2}
+i\dot\Phi_n\left(t_1\right)-i\dot\Phi_n\left(t_2\right)\right)G_n(t_1,t_2)
\hspace{0cm}
\nonumber\\
=-(g_n+g_{n+1})G_n(t_1,t_2)+g_nG_{n-1}(t_1,t_2)
\nonumber\\
+\,g_{n+1}G_{n+1}(t_1,t_2),
\label{kin}
\end{eqnarray}
where we defined $\dot\Phi_n(t)=\sum_{j=1}^n eV_j(t).$ This
kinetic equation holds for arbitrary time dependent voltages. As
before, the boundary conditions to this equation read
\begin{eqnarray}
G_0(t_1,t_2)=-iT{\rm Re}\frac{1}{\sinh\pi T(t_1-t_2+i\epsilon)},
\hspace{1.78cm}
\nonumber\\
G_N(t_1,t_2)=-iT{\rm e}^{-i eV(t_1-t_2)}{\rm Re}\frac{1}{\sinh\pi T(t_1-t_2+i\epsilon)}.
\end{eqnarray}

\subsection{Balancing fluctuating charges and voltages}

In order to complete our simple analysis we formulate the standard
circuit theory equations, which allow to include the effect of
charge accumulation in quantum dots. Let us define the fluctuating
excess charge in the $n-$th dot $q_n.$ We assume that all
quantum dots are well described by the capacitance model, in which
case one finds
\begin{equation}
q_n=C_{n+1}V_{n+1}-C_{n}V_n-C_{gn}\sum_{j=1}^n V_j.
\label{Q}
\end{equation}
Here $C_n$ is the capacitance of the $n-$th barrier and $C_{gn}$ is
the capacitance of the $n-$th dot to the ground. The current
$I_n$ is formed by the sum of three different terms,
namely the standard Ohmic term $V_n/R_n,$ the ``kinetic'' term
$\Gamma_{n,n-1}q_{n-1}-\Gamma_{n-1,n}q_n$ and the noise term
$\xi_n$:
\begin{equation}
I_n=\frac{V_n}{R_n}+\Gamma_{n,n-1}q_{n-1}-\Gamma_{n-1,n}q_n+\xi_n.
\label{In}
\end{equation}
The variation of
the charge $q_n$ is in turn determined by the currents flowing through the
$n-$th and $n+1-$th barriers.  We get
\begin{eqnarray}
\dot q_n=I_n-I_{n+1}.
\label{dotQ}
\end{eqnarray}
Finally, the sum of all fluctuating voltages $V_n$ should be equal
to the total applied voltage,
\begin{equation}
\sum_{n=1}^NV_n=V. \label{V}
\end{equation}

Eqs. (\ref{noise3},\ref{kin},\ref{Q}-\ref{V}) form a complete set
of equations, which allow to find the first order interaction
correction to the $I-V$ characteristics for an array of quantum
dots. These equations represent a straightforward generalization
of the Langevin approach employed in Ref. \onlinecite{many3}. In
contrast to the latter, however, our present analysis accounts for the
electron dwell time in quantum dots and also non-perturbatively
treats electron transport through the scatterers. In the limit of
long dwell times $\Gamma_{n\pm 1,n} \ll 1/\tau_{\rm in}$ and small
channel transmissions $\beta_n \to 1$ (i.e. for tunnel junction
arrays) our equations are replaced by those of Ref.
\onlinecite{many3}.

\subsection{Interaction correction and shot noise}

Finally, let us establish an important relation between interaction
correction to the current and the shot noise. Performing summation
of Eqs. (\ref{In}) with the weights $R_n$, we obtain
\begin{equation}
I=\frac{V}{R_\Sigma}+\sum_{n=1}^N\frac{R_n\langle\xi_n\rangle}{R_\Sigma}.
\label{I}
\end{equation}
This
formula generalizes our previous results derived for a
coherent scatterer \cite{GZ00} and a quantum dot
\cite{GZ03} to the case of quantum dot arrays and spatially extended
disordered conductors. Eq. (\ref{I}) demonstrates that the
interaction correction to the $I-V$ curve of an array of
scatterers scales linearly with the current noise produced by
these scatterers. In the absence of noise the interaction 
correction is identically zero, and the
standard Ohm's law is recovered. Eq. (\ref{I}) will be extensively
used in our subsequent calculation.

\section{Rigorous derivation}

The phenomenological analysis presented in the previous section
clearly illustrates the relation between shot noise and
interaction effects in electron transport. Now we will demonstrate
that Eqs. (\ref{noise3},\ref{kin},\ref{dotQ}) can also be derived
within the framework of a rigorous quantum mechanical procedure.
This derivation will also allow to determine the validity range of
our Langevin approach.

We first note that for a particular case of two scatterers the
above equations follow from the effective action analysis
\cite{GZ03} after averaging of the action over mesoscopic
fluctuations. Below we will see that for the case $N=2$ these
equations yield exactly the same results as those derived in Ref.
\onlinecite{GZ03} for chaotic quantum dots. Direct generalization
of the method \cite{GZ03} to the case of quantum dot arrays,
though technically possible, turns out to be rather involved since
one should first establish the full quantum mechanical action for
the whole array and then perform its averaging over mesoscopic
fluctuations. In this case it appears more convenient to average
the action already  at the first stage of the calculation. In
order to accomplish this goal we will employ the non-linear
$\sigma-$model-type of approach combined with the Keldysh technique.
This method was proposed in Ref. \onlinecite{KA} and recently applied 
to chaotic quantum dots in Ref. \onlinecite{BN}. Below we will extend this
technique to arrays of quantum dots.

\subsection{Effective action}

In the presence of electron-electron interactions general quantum
mechanical description of both compact scatterers
\cite{SZ,GZ00,GGZ02,Naz2,BN} and extended disordered conductors
\cite{GZ,KA} can be formulated in terms of the effective action
which depends on the fluctuating Hubbard-Stratonovich fields $V_1$
and $V_2$ defined on the two branches of the Keldysh contour. In
the situation considered here the action also depends on the fluctuating Green
function $\check Q_n$ which is $2\times 2$ matrix in Keldysh space
satisfying the normalization condition
\begin{equation}
\check Q^2(t_1,t_2)=\int dt_3 \check Q(t_1,t_3)\check Q(t_3,t_2)=
\delta(t_1-t_2)\check 1, \label{norm}
\end{equation}
and on the fluctuating phases of the dots
\begin{equation}
\check\Phi_n=\Phi_n\check 1 +\frac{\Phi^-}{2}\check\sigma_z,
\end{equation}
where we defined
\begin{eqnarray}
\Phi_n=\sum_{j=1}^n\int^t_{t_0}dt' \,
e(V_{j,1}(t')+V_{j,2}(t'))/2 ,\\
\Phi^-_n=\sum_{j=1}^n\int^t_{t_0}dt' \,
e(V_{j,1}(t')-V_{j,2}(t')).
\end{eqnarray}
Here and below $\check\sigma_{x,y,z}$ are the Pauli matrices in Keldysh space.

The complete expression for the effective action of the array reads
\begin{eqnarray}
iS&=&i\sum_{n=1}^N\int_0^t dt'\frac{C_n\dot\varphi_{n}\dot\varphi^-_n}{e^2}
\nonumber\\ &&
+\,i\sum_{n=1}^{N-1}\int_0^t dt'\left(\frac{C_{gn}}{e^2}+\frac{2}{\delta_n}\right)\dot\Phi_n\dot\Phi^-_n
\nonumber\\ &&
+\,\sum_{n=1}^N\frac{1}{2}{\rm Tr}\ln\left[1+\frac{T_k^{(n)}}{4}\big(\{\check Q_{n-1},\check Q_n\}-2\big)\right]
\nonumber\\ &&
-\,\sum_{n=1}^{N-1}\frac{2\pi i}{\delta_n}{\rm Tr}\left(\left(i\frac{\partial}{\partial t}
-{\dot \Phi}_n-\frac{\dot\Phi^-_n}{2}\check\sigma_z\right)\check Q_n\right).
\label{action1}
\end{eqnarray}
Here the trace includes the summation over the channel index $k$ while
the superscript $n$ indicates the scatterer number. The boundary conditions
for the operators $\check Q$  are\cite{BN}
\begin{eqnarray}
\check
Q_0(t_1,t_2)&=&\frac{-iT(\check\sigma_z+i\check\sigma_y)}{\sinh\pi
T(t_1-t_2)} -\delta(t_1-t_2)\check\sigma_x,
\nonumber\\
\check Q_N(t_1,t_2)&=&\frac{-iT\,{\rm
e}^{-ieV(t_1-t_2)}(\check\sigma_z+i\check\sigma_y)}{\sinh\pi
T(t_1-t_2)} \nonumber\\ && -\,\delta(t_1-t_2)\check\sigma_x.
\label{boundary}
\end{eqnarray}

Physical observables can be evaluated by means of the following equation:
\begin{equation}
\langle \hat A\rangle = \int_{\check Q^2_n=\check 1}{\cal D}\check
Q_n\int{\cal D}\Phi{\cal D}\Phi^- \, A\left(\frac{\delta
}{\delta\Phi},\frac{\delta }{\delta\Phi^-}\right)\,{\rm e}^{iS}.
\label{av}
\end{equation}

In the FCS-type of approach \cite{BN,Naz2} the matrix $\check Q_n$
has the form
\begin{equation}
\check Q_n={\rm e}^{i\chi_n\check\sigma_z/2}
 \left[\hat G_n(\check\sigma_z+i\check\sigma_y)-\check\sigma_x\right]
{\rm e}^{-i\chi_n\check\sigma_z/2},
\label{QFCS}
\end{equation}
where $\chi_n$ is the time and space independent ``counting
field'' for the $n$-th quantum dot. In our problem $\chi_n$ has to
be replaced by an arbitrary fluctuating Hermitian operator.
This observation suggests the following parametrization of the operator $\check
Q_n$:
\begin{equation}
\check Q_n= {\rm e}^{i\hat W_n\check\sigma_z}
 \left[\hat G_n(\check\sigma_z+i\check\sigma_y)-\check\sigma_x\right]
{\rm e}^{-i\hat W_n\check\sigma_z},
\label{param}
\end{equation}
where $\hat G_n$ and $\hat W_n$ are Hermitian operators. Here $\hat G_n$ 
accounts for fluctuations of the electron
distribution function in the $n-$th quantum dot or, more generally,
for fluctuations of the Keldysh-Usadel Green function. The
operator $\hat W_n$ describes ``quantum'' fluctuations of the
field $\check Q_n$. It is possible to demonstrate that an
arbitrary operator $\check Q_n$ satisfying the normalization
condition (\ref{norm}) and being sufficiently close to the
``classical'' one, $\hat
G_n^{(0)}(\check\sigma_z+i\check\sigma_y)-\check\sigma_x$, can be
written in the form (\ref{param}). We further note that the
parametrization (\ref{param}) is not identical to that proposed in
Ref. \onlinecite{KA}.

Let us expand the action (\ref{action1})  to the second order
in the small operators $\hat W_n$. Then we obtain
\begin{eqnarray}
iS&=&i\sum_{n=1}^N\int_0^t dt'\frac{C_n\dot\varphi_n\dot\varphi^-_n}{e^2}
\nonumber\\ &&
+\,i\sum_{n=1}^{N-1}\int_0^t dt'\left(\frac{C_{gn}}{e^2}+\frac{2}{\delta_n} \right) \dot\Phi_n\dot\Phi^-_n
\nonumber\\ &&
+\sum_{n=1}^{N-1}\frac{2\pi}{\delta_n}{\rm Tr}\left( i{\dot \Phi}^-_n\hat G_n
-2\left[i\frac{\partial}{\partial t}
-\dot \Phi_n,\hat G_n\right]\hat W_n\right)
\nonumber\\ &&
+\sum_{n=1}^N g_n {\rm Tr}\left(
-i(\hat G_n-\hat G_{n-1})\hat w_n
+\beta_n \hat G_{n-1}\hat w_n\hat G_{n}\hat w_n
\right.
\nonumber\\ &&
\left.
+\,\frac{1-\beta_n}{2}\left[(\hat G_{n-1}\hat w_n)^2
+(\hat G_{n}\hat w_n)^2\right]
-\hat w_n^2
\right).
\nonumber
\end{eqnarray}
where $\hat w_n=\hat W_n-\hat W_{n-1}.$ The
quadratic in $\hat w_n$  terms can be decoupled with the aid of the Hubbard-Stratonovich
transformation. One finds
\begin{eqnarray}
\exp\bigg\{\sum_{n=1}^N g_n{\rm Tr}\bigg(\beta_n\hat G_{n-1}\hat w_n\hat G_{n}\hat w_n
\hspace{2.5cm}
\nonumber\\
+\frac{1-\beta_n}{2}\left[(\hat G_{n-1}\hat w_n)^2
+(\hat G_{n}\hat w_n)^2\right]
-\hat w_n^2
\bigg)\bigg\}
=
\nonumber\\
\left\langle
{\rm e}^{-i\sum_{n=1}^N {\rm Tr}(\hat\zeta_n\hat w_n)}\right\rangle_{\hat\zeta_n},
\end{eqnarray}
where we introduced the Gaussian stochastic operator $\hat
\zeta_n$ with the pair correlator
\begin{eqnarray}
\langle\zeta_n(t_1,t_2)\zeta_n(t_3,t_4)\rangle=
2g_n\delta(t_3-t_2)\delta(t_1-t_4)
\nonumber\\
-\,g_n\beta_n\big[G_{n-1}(t_3,t_2)
G_n(t_1,t_4)
\nonumber\\
+\,G_{n}(t_3,t_2)G_{n-1}(t_1,t_4)\big]
\nonumber\\
-\,g_n(1-\beta_n)\big[G_n(t_3,t_2)G_n(t_1,t_4)
\nonumber\\
+\,G_{n-1}(t_3,t_2)G_{n-1}(t_1,t_4) \big].
\label{corr}
\end{eqnarray}

\subsection{Kinetic equation and Boltzmann-Langevin approach}

We are now in a position to derive the equation of motion for the
matrix $\hat G_n$. In the metallic limit $g_n \gg 1$ it is
sufficient to restrict our analysis to the least action condition
$\delta S/\delta \hat W_n=0$ which yields
\begin{eqnarray}
\frac{4\pi}{\delta_n}\left[\frac{\partial}{\partial
t}+i\dot\Phi_n(t),\hat G_n \right] =g_n\hat G_{n-1}+g_{n+1}\hat
G_{n+1}
\nonumber\\
-\,(g_n+g_{n+1})\hat G_n+ \hat\zeta_n-\hat\zeta_{n+1}.
\label{kin5}
\end{eqnarray}
Dropping the operator notations one can rewrite the same equation
in the form
\begin{eqnarray}
\frac{4\pi}{\delta_n}\left(\frac{\partial}{\partial
t_1}+\frac{\partial}{\partial t_2}
+i\dot\Phi_n(t_1)-i\dot\Phi_n(t_2)\right)G_n=g_nG_{n-1}
\nonumber\\
+\,g_{n+1}G_{n+1}-(g_n+g_{n+1})G_n +\zeta_n-\zeta_{n+1}.
\label{kin4}
\end{eqnarray}
Under the condition $g_n \gg 1$ one can also neglect the noise terms
$\zeta_{n}$ in Eq. (\ref{kin4}) which can only contribute in
higher orders in $1/g_n$. Dropping these terms one observes the
equivalence of the equations (\ref{kin4}) and (\ref{kin}).

It is also useful to illustrate a simple relation between  our Eq.
(\ref{kin4}) and the standard Boltzmann-Langevin approach
\cite{bb}, which is frequently used, e.g., for the analysis of the
shot noise in disordered conductors. Let us again define $t=(t_1+t_2)/2$
and $s=t_1-t_2$.  We will assume that $G_n$
varies slowly with $t$ and $\dot\Phi_n$ is a slow function of
time. Replacing $i\dot\Phi_n(t_1)-i\dot\Phi_n(t_2)\to
i\ddot\Phi_n(t) s$ and performing the Fourier transformation of
Eq. (\ref{kin4}) with respect to $s$, one arrives at
the equation for the distribution function
\begin{eqnarray}
\frac{4\pi}{\delta_n}\left(\frac{\partial f_n}{\partial
t}+\ddot\Phi_n\frac{\partial f_n}{\partial E}\right)=
g_nf_{n-1}+g_{n+1}f_{n+1}
\nonumber\\
-(g_n+g_{n+1})f_n -\eta_n + \eta_{n+1}, \label{BL1}
\end{eqnarray}
where $2\eta_n=\int ds\,{\rm e}^{iEs}\,\zeta_n(t+s/2,t-s/2)$ and
\begin{eqnarray}
\langle\eta_{n}(t,E)\eta_m(t',E')\rangle=2\pi
g_n\delta_{mn}\delta(t-t')\delta(E-E') \nonumber\\ \times \big\{
\beta_n[f_n(t,E)h_{n-1}(t,E)
+f_{n-1}(t,E)h_{n}(t,E)]
\nonumber\\
(1-\beta_n)[f_n(t,E)h_{n}(t,E) 
+f_{n-1}(t,E)h_{n-1}(t,E)] \big\}.\label{BL2}
\end{eqnarray}
Eqs. (\ref{BL1}), (\ref{BL2}) represent an extension of the
Boltzmann-Langevin approach \cite{bb} to arrays of quantum dots.
Within the framework of the adiabatic approximation the
latter approach follows directly from our rigorous analysis.

\subsection{Excess charges}

Let us now integrate out the field $\Phi^-_n$. Being performed
with the action (\ref{action1}) this integral gives the functional
delta-function equivalent to the equation $\delta S/\delta
\Phi^-_n=0$. This equation yields
\begin{eqnarray}
-\left(C_{gn}+\frac{2e^2}{\delta_n}\right)\sum_{j=1}^n\dot V_j-C_n\dot V_n +C_{n+1}\dot V_{n+1}
\nonumber\\
-\frac{2\pi e}{\delta_n}\frac{d}{d t}\, G_n(t,t)
=0. \label{C2}
\end{eqnarray}
The last term in Eq. (\ref{C2}) can be expressed via the
excess charge $q_n$ in the $n-$th dot. In the stationary case this charge
variable is defined as follows
\begin{eqnarray}
q_n=-\frac{2e}{\delta_n}\int
dE\big(f_n(E+\dot\Phi_n)-f_F(E)\big)
\hspace{1.8cm}
\nonumber\\
=\frac{2\pi e}{\delta_n}\lim_{s\to
0}\bigg(G_n\left(t+\frac{s}{2},t-\frac{s}{2}\right)
+\,\frac{i}{\pi s}\bigg)+\frac{2e\dot\Phi_n}{\delta_n}. \label{Qn}
\end{eqnarray}
This definition can be applied to non-stationary situations as
well. Then Eq. (\ref{C2}) is replaced by Eq. (\ref{Q}).

Finally, let us consider the limit $t_1\to t_2$ in Eq. (\ref{kin4}).
Setting $t=(t_1+t_2)/2,$ we arrive at Eqs. (\ref{In},\ref{dotQ}) where the
noise terms $\xi_n$ are identified as  $\xi_n(t)=e\zeta(t,t)/2.$ With
the aid of Eq. (\ref{corr}) one can check that the noise correlator
$\langle\xi_n(t_1)\xi_n(t_2)\rangle$ is given by the formula
(\ref{noise3}). This observation completes our derivation.

\section{Shot noise}
The first and immediate application of our formalism concerns the
analysis of the shot noise in arrays of scatterers and/or quantum
dots in the absence of interactions. We will employ Eqs.
(\ref{noise2},\ref{kin1},\ref{In}) and evaluate the noise spectrum of 
quantum dot arrays in the zero frequency limit. Our
procedure is similar to that applied in Ref. \onlinecite{Sukh} to
arrays of identical chaotic cavities.

From
Eqs. (\ref{kin1},\ref{boundary1}) we obtain
\begin{eqnarray}
f_n(E)=(1-a_n)f_F(E)+a_nf_F(E-eV), \label{fn}
\end{eqnarray}
where we have defined $a_n=\sum_{j=1}^n R_j/R_\Sigma$.
Substituting the result (\ref{fn}) into Eq. (\ref{noise2}) we
derive the noise spectrum for the $n-$th junction.
In the zero frequency limit one finds
\begin{eqnarray}
{\cal
S}_n=2\langle|\xi_n|_{\omega=0}^2\rangle =\big[1-a_n(1-a_n)-a_{n-1}(1-a_{n-1})
\nonumber\\  
-\,\beta_n(a_n-a_{n-1})^2\big]\frac{4T}{R_n} +\big[a_{n-1}(1-a_{n-1}) 
\nonumber\\
+\,a_n(1-a_n)+\beta_n(a_n-a_{n-1})^2\big]\frac{2eV}{R_n}\coth\frac{eV}{2T}.
\hspace{0.2cm}
\label{Sn}
\end{eqnarray}
Finally, we note that at sufficiently low frequencies the term
with charges $q_n$ in the right-hand side of Eq. (\ref{dotQ}) can
be neglected. In this limit the current fluctuations $\delta I$ in
the whole array are related to the current and voltage
fluctuations across the $n$-th scatterer as
\begin{equation}
\delta I=\frac{\delta V_n}{R_n}+\xi_n,
\end{equation}
where fluctuating voltages are subject to the constraint
$\sum_{n=1}^N\delta V_n=0.$ We obtain $ \delta I=(1/R_\Sigma
)\sum_{n=1}^N R_n\xi_n$ and
$$
{\cal S}={\sum_{n=1}^N R_n^2{\cal S}_n}/{R_\Sigma^2}.
$$
Then with the aid of Eq. (\ref{Sn}) we get
\begin{equation}
{\cal S}=(1-\tilde \beta)\,\frac{4T}{R_\Sigma}+\tilde
\beta\,\frac{2eV}{R_\Sigma}\coth\frac{eV}{2T},
\end{equation}
where $\tilde \beta$ is the Fano factor for 1D arrays of chaotic
quantum dots, which is obtained in the form
\begin{equation}
\tilde \beta
=\frac{1}{3}+\sum_{n=1}^N\frac{R_n^3}{R_\Sigma^3}\left(\beta_n-\frac{1}{3}\right).
\label{Farb}
\end{equation}
For arrays of diffusive scatterers with $\beta_n=1/3$ one
obviously gets $\tilde \beta =1/3$ for all values $R_n$. For
homogeneous arrays with $R_n=R$ and $\beta_n=\beta$ Eq.
(\ref{Farb}) yields
\begin{equation}
\tilde \beta =\frac{1}{3}+\frac{1}{N^2}\left(\beta
-\frac{1}{3}\right). \label{Fhom}
\end{equation}
This result demonstrates that in the limit $N\to \infty$ an array
of arbitrary -- not necessarily diffusive -- scatterers should
behave as a diffusive conductor with $\tilde \beta \to 1/3$.
In the case of identical transmissions for all conducting channels
Eq. (\ref{Fhom}) reduces to that derived in Ref. \onlinecite{Sukh}.

\section{General expression for the current}

Let us now turn to the calculation of the current-voltage
characteristics in the presence of electron-electron
interactions. According to Eq. (\ref{I}), in order to accomplish
this goal it is necessary to evaluate the average value for the
noise terms $\xi_n.$ This average
would vanish identically, $\langle\xi_n\rangle=0$, should there be
no dependence between the fluctuating voltage $V_n$ and noise
$\xi_n$. However, since such a dependence in general exists, the
averages $\langle\xi_n\rangle$ differ from zero and the
interaction correction remains finite.

In this paper we will restrict ourselves to the perturbation
theory and provide a general expression for the interaction
correction to the $I-V$ curve. In what follows we will consider
noise as a small perturbation, in which case in the leading
approximation fluctuations of the phase on $n-$th junction and the
noise $\xi_m$ are related to each other by means of a simple
formula
\begin{equation}
\delta\varphi_n(t)=-\frac{1}{e}\sum_{m=1}^N\int d\tau
K_{nm}(t-\tau)\xi_m(\tau).
\end{equation}
An explicit expression for the function $K_{nm}(\tau)$ will be
specified later. Now we only point out that due to the causality
requirement one has $K_{nm}(\tau <0 )=0$. Making use of the above
relation in the lowest non-vanishing order one can express the
average value of $\xi_n$ in the form
\begin{eqnarray}
\langle\xi_n(t)\rangle=\sum_{m=1}^N\left\langle \int dt_1\,
\frac{\delta\xi_n(t)}{\delta\varphi_m(t_1)}\delta\varphi_m(t_1)
\right\rangle
\hspace{2cm}
\nonumber\\
=-\frac{1}{e}\sum_{m,k=1}^N\int dt_1dt_2\,K_{mk}(t_1-t_2)
\left\langle\frac{\delta\xi_n(t)}{\delta\varphi_m(t_1)}\xi_k(t_2)
\right\rangle. \label{xi1}
\end{eqnarray}
Here the derivative $\delta\xi_n(t)/\delta\varphi_m(t_1)$ accounts
for the feedback of the phase fluctuations on the shot noise.
Formally this effect is encoded in the Green function
$G_n(t_1,t_2)$ which satisfies Eq. (\ref{kin}) and determines the
noise correlator (\ref{noise3}). In the lowest non-vanishing order
in $\xi_n$ it is sufficient to employ Eq. (\ref{kin}) instead of
(\ref{kin4}). From the causality requirement one finds
$\delta\xi_n(t)/\delta\varphi_m(t_1)=0$ for $t_1>t$. Utilizing
this property together with a similar one for the function
$K_{mk}(t)$ and making use of the fact that the noise variables
for different scatterers are uncorrelated one can rewrite Eq.
(\ref{xi1}) in the form
\begin{eqnarray}
\langle \xi_n\rangle =-\frac{1}{e}\sum_{m=1}^N\int
dt_1dt_2\,K_{mn}(t_1-t_2)
\frac{\delta\left\langle\xi_n(t)\xi_n(t_2)
\right\rangle}{\delta\varphi_m(t_1)}.\label{xi2}
\end{eqnarray}

In order to evaluate the functional derivative in Eq. (\ref{xi2})
it is necessary to resolve Eq. (\ref{kin}). Proceeding
perturbatively in $\delta\varphi_m$ and expressing the Green
function $G_n(t_1,t_2)$ in the form
\begin{equation}
G_n(t_1,t_2)={\rm e}^{-i\Phi_n(t_1)+i\Phi_n(t_2)}U_n(t_1,t_2),
\end{equation}
we arrive at the equation for the function $U_n(t_1,t_2)$
\begin{eqnarray}
\frac{4\pi}{\delta_n}\left(\frac{\partial}{\partial
t_1}+\frac{\partial}{\partial t_2}\right)U_n(t_1,t_2)=
-(g_n+g_{n+1})U_n(t_1,t_2)
\nonumber\\
+\,g_n{\rm e}^{i\varphi_n(t_1)-i\varphi_n(t_2)}U_{n-1}(t_1,t_2)
\nonumber\\
+g_{n+1}{\rm
e}^{-i\varphi_{n+1}(t_1)+i\varphi_{n+1}(t_2)}U_{n+1}(t_1,t_2).
\label{U}
\end{eqnarray}
With the aid of this equation we obtain
\begin{eqnarray}
\delta U_n(t_1,t_2)=\frac{i\delta_n}{4\pi}{\rm
e}^{i\Phi_n(t_1)-i\Phi_n(t_2)}\sum_{m=1}^{N}\int
d\tau\,g_m\delta\varphi_m(\tau)
\nonumber\\
\times\,\big\{
\big(D_{nm}(t_1-\tau)-D_{nm}(t_2-\tau)\big)G_{m-1}^{(0)}(t_1-t_2)
\nonumber\\
-\,\big(D_{n,m-1}(t_1-\tau)-D_{n,m-1}(t_2-\tau)\big)G_{m}^{(0)}(t_1-t_2)
\big\}.
\label{Gnpert}
\end{eqnarray}
Here $G_n^{(0)}(t_1-t_2)$ is the solution of Eq. (\ref{kin})
obtained for $\delta\dot\Phi_n=0$. It reads
\begin{equation}
G_n^{(0)}(s)=\frac{-iT}{\sinh\pi Ts}\left(1-a_n+a_n{\rm
e}^{-ieVs}\right), \label{G0}
\end{equation}
where the coefficients $a_n$ are defined after Eq. (\ref{fn}) and
the ``diffuson'' $D_{nm}(t)$ satisfies the equation
\begin{eqnarray}
\frac{\partial D_{n,m}}{\partial
t}&=&\frac{\delta_n}{4\pi}\big(g_nD_{n-1,m}+g_{n+1}D_{n+1,m}
\nonumber\\ && -(g_n+g_{n+1})D_{n,m}\big)+\delta_{nm}\delta(t)
\label{Deq}
\end{eqnarray}
with the boundary conditions
\begin{equation}
D_{0,m}=D_{N,m}=D_{n,0}=D_{n,N}=0. \label{Dbound}
\end{equation}
As before, due to causality one has $D_{nm}(t)=0$ for $t<0.$

With the aid of the above expressions we get
\begin{eqnarray}
\frac{\delta G_n(t,t_2)}{\delta\varphi_m(t_1)}=
-iG_n^{(0)}(t-t_2)
\hspace{3.8cm}
\nonumber\\
\times \big[\delta(t-t_1)-\delta(t_2-t_1)\big]
\theta(n-m+0)
\hspace{2.3cm}
\nonumber\\
+\frac{i\delta_ng_m}{4\pi}\big\{
\big[D_{nm}(t-t_1)-D_{nm}(t_2-t_1)\big]G_{m-1}^{(0)}(t-t_2)
\nonumber\\
-\big[D_{n,m-1}(t-t_1)-D_{n,m-1}(t_2-t_1)\big]G_{m}^{(0)}(t-t_2)
\big\}.
\end{eqnarray}
This equation in combination with Eqs. (\ref{noise3},\ref{xi2}) enables one
to evaluate the derivative ${\delta\left\langle\xi_n(t)\xi_n(t_2)
\right\rangle}/{\delta\varphi_m(t_1)}$ and derive the final
expression for the current. With the aid of Eq. (\ref{I}) we obtain
\begin{equation}
I=\frac{V}{R_\Sigma}+\delta I, \label{If}
\end{equation}
where $\delta I$ is the interaction correction which can be split
into two parts $\delta I=\delta I_1+\delta I_2$:
\begin{eqnarray}
\delta I_1&=&\frac{1}{4\pi^2 eR_\Sigma}\sum_{n,m=1}^N g_m \int
dtdt'\frac{\pi^2T^2\sin eVt}{\sinh^2\pi Tt} \nonumber\\ &&\times\,
K_{mn}(t-t') \big[\delta_{n-1}(a_{m-1}-a_{n-1})D_{n-1,m}(t')
\nonumber\\ && -\,\delta_{n-1}(a_{m}-a_{n-1})D_{n-1,m-1}(t')
\nonumber\\&& +\,\delta_{n}(a_{m-1}-a_{n})D_{n,m}(t')
\nonumber\\&& -\, \delta_{n}(a_{m}-a_{n})D_{n,m-1}(t') \big],
\label{dI2}
\end{eqnarray}
\begin{eqnarray}
\delta I_2=-\frac{1}{\pi eR_\Sigma^2}\sum_{n,m=1}^N \beta_nR_n\int
dtdt'\frac{\pi^2T^2\sin eVt}{\sinh^2\pi Tt} \nonumber\\ \times\,
K_{mn}(t-t')\bigg[\delta_{nm}\delta(t')
-\frac{\delta_{n-1}g_m}{4\pi}\big(D_{n-1,m-1}(t')
\nonumber\\
-\,D_{n-1,m}(t')\big)-\frac{\delta_{n}g_m}{4\pi}\big(D_{nm}(t')
-D_{n,m-1}(t')\big) \bigg]. \label{dI1}
\end{eqnarray}
Eqs. (\ref{If}-\ref{dI1}) represent our general result for the
$I-V$ curve of a 1d array of metallic quantum dots in the presence
of interactions. One can verify that in the particular case of two
scatterers or, equivalently, for a single chaotic quantum dot,
Eqs. (\ref{If}-\ref{dI1}) reduce to the expressions derived in
Ref. \onlinecite{GZ03} by means of a different approach. Let us
also note that the expression for the function $K_{mn}(t)$ is
determined by the solution of Eqs. (\ref{Q}-\ref{dotQ}) under the
constraint (\ref{V}). Below we will explicitly find this solution
for the specific case of homogeneous arrays of quantum dots.

\section{Homogeneous 1D array}

Consider an array formed by the scatterers and quantum dots with
identical parameters. In what follows we set $C_n=C,$
$C_{gn}=C_g,$ $g_n=g,$ $\beta_n=\beta,$ $R_n=R,$
$\delta_n=\delta.$ In this case it is straightforward to derive
the exact expressions for the functions $K_{mn}(t)$ and $D_{nm}$.
These expressions read
\begin{eqnarray}
K_{mn}(t)=\frac{2e^2}{N}\sum_{q=1}^{N-1}\int\frac{d\omega}{2\pi}\,\frac{{\rm
e}^{-i\omega t}}{-i\omega+0} Z_{\omega q}
\nonumber\\
\times\,\cos\left(\frac{\pi qn}{N}-\frac{\pi q}{2N}\right)
\cos\left(\frac{\pi qm}{N}-\frac{\pi q}{2N}\right), \label{K}
\end{eqnarray}
where we defined the impedance $Z_{\omega q}$
\begin{eqnarray}
Z_{\omega q}= \frac{1}{\left(-i\omega +\frac{1-\cos\frac{\pi
q}{N}}{\tau_D}\right)\left(C+\frac{C_g}{2(1-\cos\frac{\pi
q}{N})}\right) +\frac{1}{R}}, \label{Z}
\end{eqnarray}
and
\begin{eqnarray}
D_{nm}&=&\frac{2}{N}\sum_{q=1}^{N-1}\int\frac{d\omega}{2\pi}\,{\rm
e}^{-i\omega t} D_{\omega q}\sin\frac{\pi qn}{N}\sin\frac{\pi
qm}{N},
\nonumber\\
D_{\omega q}&=&\frac{1}{-i\omega+\frac{1-\cos(\pi q/N)}{\tau_D}}.
\label{D}
\end{eqnarray}
Here and below $\tau_D=2\pi/g\delta$ stands for the electron dwell
time in a single quantum dot and the coefficients $a_n$ reduce to
$a_n=n/N$. Making use of the property $K_{mn}(t)=K_{nm}(t)$ and
$D_{mn}(t)=D_{nm}(t)$ we obtain
\begin{eqnarray}
\delta I_1&=& -e\sum_{q=1}^{N-1}\frac{1+\cos\frac{\pi
q}{N}}{2\tau_DRN^2}
\int\frac{d\omega}{2\pi}\, {\rm Im}(Z_{\omega q}D_{\omega
q}^2)\,B(\omega,V,T) \nonumber\\ &&
+\,e\sum_{q=1}^{N-1}\frac{1+\cos\frac{\pi q}{N}}{\tau_D^2RN^3}
{\rm Im}\int\frac{d\omega}{2\pi} Z_{\omega q}D_{\omega q}^3
B(\omega,V,T) \nonumber\\ &&\times\, \frac{1-(-1)^qu^N(\omega)}
{1+(-1)^qu^N(\omega)}\sqrt{-i\omega\tau_D(2-i\omega\tau_D)},
\label{dI22}
\end{eqnarray}
and
\begin{eqnarray}
\delta
I_2=-\frac{e\beta}{2N^2R}\sum_{q=1}^{N-1}\int\frac{d\omega}{2\pi}\,
{\rm Im}(Z_{\omega q}D_{\omega q})B(\omega,V,T), \label{dI11}
\end{eqnarray}
where $B(\omega,V,T)=\sum_{\pm}(eV\pm\omega)\coth\frac{\omega\pm
eV}{2T}$ and $u(\omega)=1-i\omega-\sqrt{(1-i\omega\tau_D)^2-1}$.
We observe that the second contribution to the interaction
correction $\delta I_2$ scales with the Fano factor $\beta$ of 
individual scatterers and, hence, vanishes for $\beta \to 0$. At
the same time the first contribution $\delta I_1$ does not depend on
$\beta$, i.e. it is {\it universal} for any type of scatterers. We
also note that the term $\delta I_1$ differs from zero for all
$N>2$ but vanishes identically in the case of two scatterers
$N=2$.

The frequency integrals in Eqs. (\ref{dI22}) and (\ref{dI11}) can
be performed exactly with the result
\begin{widetext}
\begin{eqnarray}
\delta I&=&-\frac{2 Te}{ N^2}\sum_{q=1}^{N-1}\bigg[\beta
+\frac{\cos\frac{\pi q}{N}}{4\sin^2\frac{\pi q}{2N}}\bigg] \bigg[
W\bigg(\frac{2\sin^2\frac{\pi q}{2N}}{\pi TR\big(4C\sin^2\frac{\pi
q}{2N}+C_g\big)} +\frac{\sin^2\frac{\pi q}{2N}}{\pi T\tau_D}
+\frac{ieV}{2\pi T}\bigg) -W\bigg(\frac{\sin^2\frac{\pi
q}{2N}}{\pi T\tau_D}+\frac{ieV}{2\pi T}\bigg)\bigg] \nonumber\\
&&
-\,\frac{4Te}{N^4}\sum_{p,q=1}^{N-1}\frac{(1-(-1)^{p+q})\sin^2\frac{\pi
q}{N}\sin^2\frac{\pi p}{N}} {\big(\cos\frac{\pi
q}{N}-\cos\frac{\pi p}{N}\big)^3 \big(2\sin^2\frac{\pi
p}{2N}+\frac{R}{2\tau_D}\big(4C\sin^2\frac{\pi
p}{2N}+C_g\big)\big(\cos\frac{\pi q}{N}-\cos\frac{\pi
p}{N}\big) \big)} \nonumber\\ && \times\, \bigg[
W\bigg(\frac{2\sin^2\frac{\pi p}{2N}}{\pi TR\big(4C\sin^2\frac{\pi
p}{2N}+C_g\big)} +\frac{\sin^2\frac{\pi p}{2N}}{\pi T\tau_D}
+\frac{ieV}{2\pi T}\bigg) -W\bigg(\frac{\sin^2\frac{\pi
q}{2N}}{\pi T\tau_D}+\frac{ieV}{2\pi T}\bigg)\bigg].
\label{dIexact}
\end{eqnarray}
\end{widetext}
Here we defined the function $W(x)={\rm Im}\,[x\Psi (1+x)]$, where $\Psi (x)$
is the digamma function. Eq. (\ref{dIexact}) is the exact expression 
for the leading (in $1/g$) interaction correction to the current 
valid both in linear and non-linear in voltage regimes and for any 
number of scatterers $N$ in the system.

Let us consider a physically important limit of relatively large
metallic quantum dots with $RC,RC_g\ll\tau_D$. Making use of this
inequality one can significantly simplify the general result
(\ref{dIexact}) and find
\begin{eqnarray}
\delta I= -\frac{2Te}{
N^2}\sum_{q=1}^{N-1}
\bigg[\beta
W\bigg(\frac{2\sin^2\frac{\pi
q}{2N}}{\pi TR\left(4C\sin^2\frac{\pi q}{2N}+C_g\right)}
+\frac{ieV}{2\pi T}\bigg)
\nonumber\\
-\bigg[ \beta -
\frac{1-\frac{2(1-(-1)^q)}{N^2}\cot^2\frac{\pi
q}{2N}}{1-\cos\frac{\pi q}{N}}\bigg] W\bigg(\frac{\sin^2\frac{\pi
q}{2N}}{\pi T\tau_D}+\frac{ieV}{2\pi T}\bigg)\bigg].
\hspace{0.3cm}
\label{dIapp}
\end{eqnarray}
In the linear in voltage regime the above expression yields the
result for the zero bias conductance of the array $G=1/NR+\delta
G$. The interaction correction $\delta G$ takes the form
\begin{eqnarray}
\delta G= -\frac{e^2}{\pi N^2}\sum_{q=1}^{N-1}\bigg[\beta
L\bigg(\frac{2\sin^2\frac{\pi q}{2N}}{\pi
TR\left(4C\sin^2\frac{\pi q}{2N}+C_g\right)} \bigg)
\nonumber\\
 -\bigg(\beta -
\frac{1-\frac{2(1-(-1)^q)}{N^2}\cot^2\frac{\pi
q}{2N}}{1-\cos\frac{\pi q}{N}}\bigg) L\bigg(\frac{\sin^2\frac{\pi
q}{2N}}{\pi T\tau_D}\bigg)\bigg],
\label{dGapp}
\end{eqnarray}
where we have defined $L(x)=\Psi(1+x)+x\Psi'(1+x).$

\begin{figure}
\begin{center}
\includegraphics[width=8.5cm]{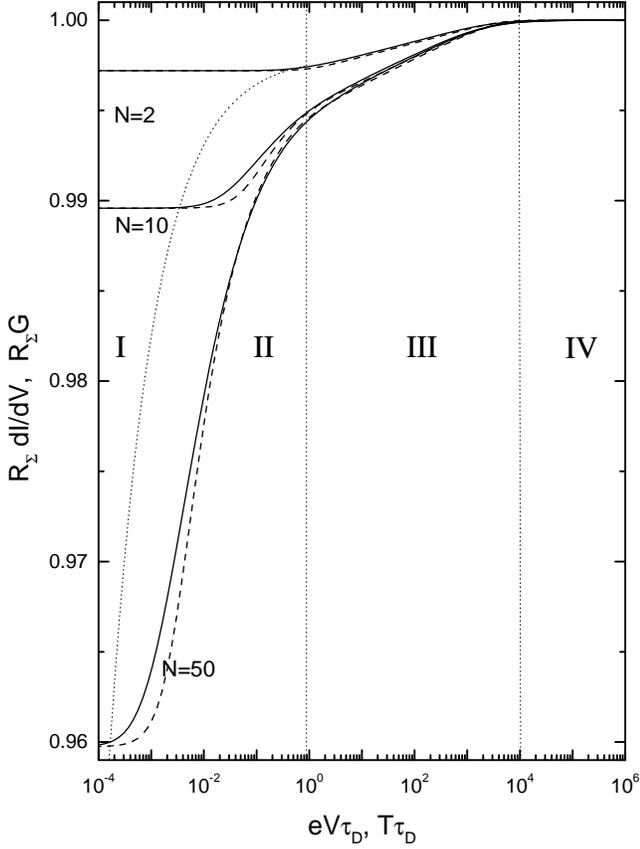}
\end{center}
\caption{The linear conductance $G$ (solid lines) as a function of temperature
$T$ together with the differential conductance $dI/dV$ (dashed lines) as a
function of the applied voltage $V$ at $T=0$. The results are
obtained from Eq. (\ref{dIexact})
for $g=1000,$ $\beta=1/3,$ $\tau_D/RC=10^4,$ $C_g/C=2.5$
and for three different numbers of barriers: $N=2,10$ and 50. We
identify four different regimes (the boundaries between them are
shown by dotted lines): (I) saturation regime $eV,T\lesssim
\pi^2/2N^2\tau_D $, Eq. (\ref{G00}), (II) diffusive regime
$\pi^2/2N^2\tau_D\lesssim eV,T\lesssim 1/\tau_D$, Eqs.
(\ref{G001},\ref{IV001}), (III) logarithmic regime of almost
independent barriers $1/\tau_D \lesssim eV,T\lesssim 1/RC$, Eqs.
(\ref{G00L},\ref{IVL}) and (IV) high temperature (classical)
regime $eV,T\gtrsim 1/RC$, Eq. (\ref{Gcl}).}
\end{figure}

Let us now briefly analyze the above results in
various limits. The case $N=2$ was already considered in details
in Ref. \onlinecite{GZ03}, here we will concentrate on the behavior
of quantum dot arrays containing many scatterers $N\gg 1.$ 
In this case the expression
for the interaction correction (\ref{dGapp}) can be further
simplified. In the high temperature limit $TRC_m\gg 1$ (where
$C_m=$min$[C,C_g]$) we obtain
\begin{equation}
\delta G=-\frac{e^2\beta
}{6NTRC}\bigg(1-\sqrt{\frac{C_g}{C_g+4C}}\bigg).
\label{Gcl}
\end{equation}
At intermediate temperatures $1/RC,1/RC_g \gg T \gg 1/\tau_D$ Eq. (\ref{dGapp}) yields
\begin{eqnarray}
\delta G\approx-\frac{e^2\beta }{\pi N}\left[\ln\frac{1}{2\pi TR
C^*}+1\right],
\label{G00L}
\end{eqnarray}
where $C^*=\left(\sqrt{C_g}+\sqrt{C_g+4C}\right)^2/4.$ In the
interval $\pi^2/2N^2\tau_D\ll T\ll 1/\tau_D$ we find
\begin{eqnarray}
\delta G\approx-\frac{e^2\beta}{\pi
N}\ln\frac{2\tau_D}{RC^*}-\frac{e^2}{\pi
N}\frac{3\zeta(3/2)}{4\sqrt{\pi T\tau_D}} \hspace{2.5cm}
\nonumber\\
+\,\frac{e^2}{\pi N}\left[1+\left(\beta
-\frac{1}{4}\right)\left(1-\frac{\gamma}{2}\right)\sqrt{\pi
T\tau_D} +\frac{\pi}{3NT\tau_D} \right], \label{G001}
\end{eqnarray}
where $\gamma \simeq 0.577$ is the Euler constant. Finally, in the
limit of very low temperatures, $T\ll \pi^2/2N^2\tau_D,$ the
correction to the conductance saturates, and we get
\begin{eqnarray}
\delta G\approx-\frac{e^2\beta}{\pi
N}\ln\frac{2\tau_D}{RC^*}-\left(0.368-\frac{1}{N}\right)
\frac{e^2}{\pi}. \label{G00}
\end{eqnarray}

Consider now the nonlinear regime $e|V|\gg T$ in which case the
$I-V$ curve is determined by Eq. (\ref{dIexact}) (or
(\ref{dIapp})). We will again consider the limit $N \gg 1$ and
make use of Eq. (\ref{dIapp}). At very small voltages and
temperatures, $e|V|\ll \pi^2/2N^2\tau_D,$ one finds $dI/dV=1/NR+\delta
G$, where $\delta G$ is again defined by Eq. (\ref{G00}). At
higher voltages, $\pi^2/2N^2\tau_D \ll e|V|\ll 1/\tau_D$ we obtain
\begin{eqnarray}
\frac{dI}{dV}=\frac{1}{NR}-\frac{e^2\beta}{\pi
N}\ln\frac{2\tau_D}{RC^*}-\frac{e^2}{\pi
N}\frac{1}{\sqrt{2e|V|\tau_D}}.
\label{IV001}
\end{eqnarray}
At even higher voltage, $1/\tau_D\ll e|V|\ll 1/RC^*$, the
differential conductance takes the form
\begin{equation}
\frac{dI}{dV}=\frac{1}{NR}-\frac{e^2\beta}{\pi
N}\ln\frac{1}{e|V|RC^*}.
\label{IVL}
\end{equation}

The linear conductance $G$ as a function of temperature and the
differential conductance $dI/dV$ at $T \to 0$ as a function of the
applied voltage are depicted in Fig. 2 for different number of
scatterers $N$ in the system. One observes that both quantities
(as functions of corresponding variables) demonstrate a very
similar behavior. In both cases four different regimes can be
distinguished, as it was already specified above. Further
discussion of these results is postponed to Sec. VIII.

\section{Effect of environment}

In order to complete our analysis let us also include the effect
of external leads into consideration. For simplicity we will
follow the standard procedure and assume the leads impedance to be
purely Ohmic. This procedure amounts to replacing Eq. (\ref{V}) by
a more general one,
\begin{equation}
\sum_{j=1}^N V_j + IR_S = V_x, \label{Vnew}
\end{equation}
where $V_x$ is the voltage applied to the whole system
``array+leads'', $R_S$ is the resistance of the leads and $I=C\dot
V_n + I_n$ is the current flowing through the leads. Due to
current conservation the scatterer number $n$ can be chosen
arbitrarily here. Eq. (\ref{Vnew}) is solved together with Eqs.
(\ref{noise3},\ref{kin2},\ref{Q},\ref{In},\ref{I}) in exactly the
same way as it was done above in the absence of the shunt. As a
result, the effect of $R_S$ is accounted for by means of a simple
replacement
\begin{equation}
K_{nm}(t)\longrightarrow K_{nm}(t)+K_S(t),
\end{equation}
where $K_{nm}(t)$ is defined in Eq. (\ref{K}) and
\begin{equation}
K_S(t)=\frac{e^2}{N}\int\frac{d\omega}{2\pi}\, \frac{{\rm
e}^{-i\omega
t}}{(-i\omega+0)\left(-i\omega+\frac{1}{R}+\frac{N}{R_S}\right)}.
\label{KS}
\end{equation}
Note that the function $K_S(t)$ does not depend on $n$ and $m.$
After such a replacement the new expression for $K_{nm}(t)$ should
be substituted into Eqs. (\ref{dI2},\ref{dI1}) and we arrive at
the result
\begin{equation}
I=\frac{V}{R_\Sigma}+\delta I+\delta I_S,
\end{equation}
where $V=\sum_{j=1}^N \langle V_j\rangle$ is the average voltage
across the array and  $\delta I$ is defined in Eq.
(\ref{dIexact}). The general expression for the additional term
$\delta I_S$ is rather cumbersome and will not be presented here.
Below we will only address the effect of external leads on the
linear conductance in the low temperature limit $T <
\pi^2/2N^2\tau_D$. It turns out that for non-zero $R_S$ --
similarly to the case of single quantum dots \cite{GZ03} --
the conductance saturation is lifted and the result
(\ref{G00}) becomes incomplete. Taking into account the shunt
contribution to the current $\delta I_S$, one finds
\begin{eqnarray}
G&\approx&\frac{1}{NR}-\frac{e^2\beta}{\pi
N}\ln\frac{2\tau_D}{RC^*} -\left(0.368-\frac{1}{N}\right)
\frac{e^2}{\pi} \nonumber\\ && -\frac{e^2\tilde
\beta}{\pi}\frac{R_S}{R_S+NR} \ln\frac{\pi^2}{2N^2T\tau_D},
\label{GS}
\end{eqnarray}
where
\begin{equation}
\tilde \beta =\frac{\beta}{N^2}+\frac{1}{N^4}
\sum_{q=1}^{N-1}\frac{(1-(-1)^q)\cos^2\frac{\pi q}{2N}}
{\sin^4\frac{\pi q}{2N}}. \label{ffano}
\end{equation}
The sum in (\ref{ffano}) is evaluated exactly and just yields the
Fano factor of the array (\ref{Fhom}). Thus, in the presence of
an external shunt the conductance keeps decreasing
logarithmically with $T$ even at very low temperatures. As before,
this logarithmic contribution scales linearly with the total Fano
factor of the array $\tilde \beta$ which tends to the universal
value 1/3 in the limit of large $N$. This result is in the
agreement with our previous findings \cite{GZ00,GZ03} and once
again emphasizes a direct relation between shot noise and
interaction effects in disordered conductors.

\section{Discussion}

In this paper we have proposed a general model which embraces
virtually any type of disordered conductors and allows to account
for Coulomb interaction effects in electron transport through such
conductors. Exploiting an intimate relation between shot noise
and interaction effects, in Sec. II and III we derived a closed
set of Langevin-type of equations which allow to conveniently
study electron transport in the presence of electron-electron 
interactions \cite{FN}. The key idea of our approach is to account for
modifications of the shot noise due to non-equilibrium effects and
to self-consistently describe these effects and their impact on
fluctuating charges and voltages inside the conductor. For the
sake of definiteness here we focused our attention on quasi-1D
conductors, however one can trivially extend the whole analysis to
2D and 3D conductors as well. This generalization will be carried
out elsewhere.

The technique developed in this paper allows to obtain a general
formula for the interaction correction to the current, Eq.
(\ref{dIexact}), which contains all the results derived previously
for various quasi-0D and quasi-1D disordered conductors and
extends these results to yet unexplored regimes. At
sufficiently high energies (exceeding the inverse dwell time of a
single quantum dot $1/\tau_D$) the scatterers behave as
effectively independent ones, and one can identify two different
regimes (regimes III and IV in Fig. 2) described by Eqs.
(\ref{G00L},\ref{IVL},\ref{Gcl}). At such energies the interaction
correction scales with the Fano factor $\beta$ of individual
scatterers and in a wide interval of energies depends
logarithmically on temperature or voltage. For a special case of
tunnel barriers $\beta \to 1$ our results reduce to those derived
in Refs. \onlinecite{many3,GG}, while in the limit of ballistic
contacts with $\beta \to 0$ (or, equivalently, diffusive wires 
with point-like impurities) the interaction correction turns out to 
be negligibly small in this regime. 

At energies below $1/\tau_D$ (regime II) scatterers
located sufficiently close to each other become effectively
correlated. The number of such scatterers $N_{\rm eff}$ in one ``correlated''
segment of the array grows with
decreasing temperature (or voltage) as $N_{\rm eff} \sim
1/\sqrt{T\tau_D}$ (or $N_{\rm eff} \sim 1/\sqrt{eV\tau_D}$). In 
this regime the system can be viewed as a chain of
$\sim N/N_{\rm eff}$ segments, each of them now
playing the role of a ``new'' independent scatterer with an effective
conductance $g_{\rm eff} \sim g/N_{\rm eff} \sim g\sqrt{T\tau_D}$
(or  $\sim g\sqrt{eV\tau_D}$). Then the results \cite{many3,GG} 
can be applied again,
in the corresponding expression for the interaction correction 
one should only substitute $g_{\rm eff}$ instead of $g$. In this case
the logarithmic dependence of the interaction correction on
temperature/voltage drops out \cite{FN2} and, e.g., for the linear
conductance one finds $\delta G/G \sim -\beta_{\rm eff}/g_{\rm
  eff}$, where  $\beta_{\rm eff}$ is the Fano factor of a segment with
$N_{\rm eff}$ scatterers.
According to Eq. (\ref{Fhom}), for sufficiently large $N_{\rm eff}\gg 1$
the factor $\beta_{\rm eff}$ approaches the universal value 1/3, and we
obtain  $\delta G/G \sim -1/g\sqrt{T\tau_D}$ in agreement with the
well known result \cite{AA} and also with our rigorous formula (\ref{G001})
which -- in addition -- contains a temperature-independent
contribution $\propto \beta$ coming from high energy modes. Finally,
as $N_{\rm eff}$ approaches $N$ the system conductance either saturates
(for $R_S \to 0$, regime I) or crosses over to the low energy logarithmic 
regime (\ref{GS}) caused by additional voltage fluctuations across the array
due to non-zero external shunt resistance $R_S$.

It is also straightforward to establish a direct relation between the results
derived here and those obtained diagrammatically in the linear in voltage 
regime \cite{many4,AA}. 
By setting $N \to \infty$ and $\beta \to 1$ from Eq. (\ref{dGapp})
we reproduce the results \cite{many4} for the interaction correction
in tunnel junction arrays, while in the limit $\beta \to 0$ the latter
equation yields the standard result \cite{AA} for diffusive
wires \cite{FN3}. The same equivalence can be observed
at the level of general expressions (\ref{dI22},\ref{dI11})
considered in the limit $eV\ll T$, $N \to \infty$ and for $\tau_D \gg RC,RC_g$.
For $\beta=1$ the result \cite{many4} follows from
the sum  of two terms (\ref{dI22}) and (\ref{dI11}), 
while for $\beta=0$ the second contribution  (\ref{dI11})
vanishes identically  and the result \cite{AA} is obtained
only from the first term  (\ref{dI22}). These observations 
demonstrate that the reduction of our model to 
one for diffusive wires with point-like impurities is achieved by
setting $\beta \to 0$. In the latter case $N$ coincides with the total
number of impurities in the wire.  

At last, let us briefly summarize the applicability conditions for our results.
As it was already discussed above, our Langevin approach is justified in
the metallic limit $g_n \gg 1$. Under this condition our technique should
account for all essential processes except for subtle instanton effects which
may show up only at exponentially low energies. An obvious necessary (though
possibly not sufficient) validity
condition of our results derived in the linear in voltage regime
is $\delta G /G \ll 1$. While at high enough temperatures 
this inequality is automatically fulfilled in the metallic limit $g_n \gg 1$,
at the lowest energies/temperatures a much more stringent condition 
$g_{\Sigma}=2\pi/ e^2R_\Sigma \gg 1$ has to be satisfied. The latter
condition is inevitably violated for large number of scatterers  
$N$ in which case a non-perturbative analysis becomes necessary in the low
energy limit. This analysis is beyond the frames of the present paper. 
In the non-linear regime and at sufficiently high voltages 
the applicability range of our results can be additionally restricted by
electron heating effects which we do not address in this work. 

\vspace{0.5cm}

\centerline{\bf Acknowledgements}

We are grateful to D.A. Bagrets and S.V. Sharov for stimulating
discussions. This work is part of the Kompetenznetz ``Funktionelle
Nanostructuren'' supported by the Landestiftung
Baden-W\"urttemberg gGmbH and of the STReP ``Ultra-1D'' supported
by the EU.


\begin{thebibliography}{99}
\bibitem{FN2} In this case the combination $T\tau_D^{\rm eff}$ occurs under the
logarithm, where $\tau_D^{\rm eff}$ is the total dwell time
for $N_{\rm eff}$ scatterers. This combination
reduces to a temperature independent constant of order unity since
$\tau_D^{\rm eff} \sim 1/T$.
\bibitem{FN3} In order to observe the equivalence between our 
Eq. (\ref{dGapp}) in the limit $\beta \to 0$ and the results \cite{AA} for 
the interaction correction in quasi-1D diffusive wires it suffices to
introduce the diffusion coefficient $D=a^2/2\tau_D,$ where $a$ is 
the size of a single quantum dot.
\end{thebibliography}
\end{document}